\documentclass[a4paper,prl,10pt,notitlepage,twocolumn]{revtex4-1}

\usepackage{graphicx,color}
\usepackage[colorlinks=true,linkcolor=blue,citecolor=blue]{hyperref}
\usepackage{amssymb,amsmath,bbold,mathtools}

\begin{document}

\title{On super-Poissonian behavior of the Rosenzweig-Porter model in the non-ergodic extended regime}
\author{Richard Berkovits}
\affiliation{Department of Physics, Jack and Pearl Resnick Institute, Bar-Ilan University, Ramat-Gan 52900, Israel}

\begin{abstract}
  The Rosenzweig-Porter model has seen a resurgence in interest as it exhibits a
  non-ergodic extended phase between the ergodic extended metallic phase and the localized phase. Such a
  phase is relevant to many physical  models from the Sachdev-Ye-Kitaev model in high-energy physics and quantum
  gravity, to the interacting many-body localization in condensed matter physics and quantum computing.
  This phase is characterized by fractal behavior of the wavefunctions, and a postulated
  correlated mini-band structure
  of the energy spectrum. Here we will seek evidence for the latter in the spectrum. Since this behavior is expected
  on intermediate energy scales spectral rigidity is a natural way to tease it out. Nevertheless, due to
  the Thouless energy and ambiguities in the unfolding procedure, the results are inconclusive. On the other hand,
  by using the singular value decomposition method, clear evidence for a super-Poissonian behavior in this regime emerges,
  consistent with a picture of correlated mini-bands.
\end{abstract}


\maketitle

The Anderson metal-insulator transition continues to surprise even
after six decades \cite{anderson58}.
The canonical picture for a single particle Anderson transition
is the three dimensional Anderson model which shows a metal-insulator
transition for a critical value of on-site disorder. For weak disorder the
system is metallic and the wave function is extended, while for stronger disorder the
wave function is localized and the system is insulating \cite{lee85,kramer93}. At the critical disorder
the wave function is fractal \cite{aoki83}. The energy spectrum also reflects these phases.
In the  localized regime the level spacing distribution (corresponding to small energy scales, large
times) follow the Poisson distribution, while for the extended regime
corresponds to the Wigner-Dyson (WD) distribution.
\cite{shklovskii93,ghur98,mirlin00,evers08}.
Several
forms of the level spacing were suggested at criticality  \cite{kravtsov94,evangelou94,aronov94,zharekeshev97}.
For larger energy scales, the
spectral rigidity, i.e., the variance of the number of levels in a given 
energy window is a useful indicator. For the localized phase the variance
is equal to the average number of states in this window, while for the extended
phase it is proportional to the logarithm of the average. At the
critical point the variance is proportional to the average number of states, with a proportionality
lower than one \cite{altshuler86,altshuler88,aronov95}.

An additional energy scale relevant to disordered metals
is the Thouless energy $E_T=g \Delta$ (where $g$ is the dimensionless conductance, and $\Delta$
the average level spacing)
\cite{altshuler86}. 
While WD predictions hold up to an energy scale $E<E_T$, above which a non-universal behavior takes over.
The physical origin of the
Thouless energy is the onset of diffusive behavior.

There has been a recent surge in interest in the critical behavior of the transition. Part of this
interest stems from the realization that for the many body localization phenomenon the localized
and extended regions may
be separated by a critical regime
\cite{x1,x2,x3,x4,x5,x6,x7,x8,x9,x10}.
An additional
motivation pertains to the Sachdev-Ye-Kitaev (SYK) model, originally introduced in
the study of spin
liquids \cite{sachdev93} and recently gaining relevance to
holographic dualities in string theory \cite{maldacena99} and
quantum gravity \cite{maldacena16}. There is evidence 
the SYK model perturbed by a single-body term
shows a critical region \cite{micklitz19}. This model also shows a signature of
the existence of a Thouless energy \cite{garcia16}.

The generalized Rosenzweig-Porter random matrix
model (GRP) \cite{rosenzweig60,kravtsov15}
is considered the most simple model for which
the localized and fully ergodic phases both exist, with an 
non-ergodic extended (NEE)  phase separating them.
Almost all evidence for the NEE comes from the study of the fractality
of the wave functions \cite{kravtsov15,monthus17,kravtsov18,bogomolny18,nosov19,pino19}.
In Ref. \onlinecite{detomasi19} some tantalizing clues for a super-Poissonian
behavior have appeared in the n-th level spacing distribution. 
Finding fingerprints for NEE in the energy spectrum is important, for both theoretical and
practical reasons. It it is much easier
numerically, as well as experimentally, to obtain the energies than wavefunctions for large systems.
Here, we examine two methods to garner such information. The venerable
method of spectral rigidity (number variance) \cite{mehta91},
and singular value decomposition (SVD) \cite{fossion13,torres17,torres18}.
Both will be used to study very large GRP matrices on scales of thousands of eigenvalues.
It turns out that although the spectral rigidity exhibits anomalies which could
be attributed to NEE,  it is nevertheless hard to separate them from the effects of the
Thouless energy, finite size, and dependence on unfolding. On the other hand,
SVD seems to provide strong evidence for an intermediate scale of energy,
for which systems belonging to the NEE phase show super-Poissonian behavior
of the spectra, similar to the random Cantor set behavior \cite{x2}.

The GRP is defined by a random matrix $H_{ij}$, of
size $N \times N$,
where the diagonal terms are chosen from a certain distribution while the off diagonal
term is chosen from a distribution with a variance proportional to $N^{-\gamma}$. Specifically,
we have chosen the diagonal $H_{ii}$ from a box distribution with a range
$-\sqrt{6}/2 \ldots \sqrt{6}/2$ ($\delta^2 \langle H_{ii} \rangle = 1/2$) and
the off-diagonal $H_{i\ne j}$ from a box distribution between $-N^{-\gamma/2}/2 \ldots N^{-\gamma/2}/2$
($\langle \delta^2  H_{i\ne j} \rangle = N^{-\gamma}/12$), thus corresponding to
$N^{\gamma}  \langle H_{i\ne j}^2\rangle/\langle H_{ii}^2 \rangle=1/6$.

For GRP one expects a transition from localized to extended behavior
at $\gamma=2$ \cite{kravtsov15,detomasi19,pandey95,brezin96,guhr96,altland97,kunz98,facoetti16}.
This transition shows 
in the nearest neighbor level distribution that switches from a Poisson behavior for $\gamma>2$ to
WD repulsion at $\gamma<2$.
In the supplementary material\cite{supp} we show
the finite size scaling of the ratio statistics, substantiating this transition at $\gamma=2$. 

On the other hand, the transition
between the NEE phase to the truly extended phase anticipated to occur
at $\gamma=1$, leaves no signature in the ratio statistics. This is expected, as
small energy scale correspond to long times, and since the states are extended
at very long times both in the ergodic as well as the NEE phases, short energy scales
can not resolve the difference.
Thus, one should
probe energy scales that are much larger than the mean level spacing.

Two statistical measures will be considered:
level number variance for a
given energy window 
and
applying the singular value decomposition
(SVD) on the spectrum \cite{fossion13,torres17,torres18} .

The first is also known as spectral rigidity \cite{mehta91}.  
Specifically, for an energy window of a size $E$, the average number of levels, $\langle n(E) \rangle$,
and the variance,
$\langle \delta^2 n(E) \rangle=\langle (n(E) - \langle n(E) \rangle)^2 \rangle$ are calculated.
For the WD distribution $\langle \delta^2 n(E) \rangle= 0.44+
(2/\pi^2)\ln(\langle n(E) \rangle)$, while for the Poisson distribution
$\langle \delta^2 n(E) \rangle=\langle n(E) \rangle$.
One may argue that for the
NEE phase ($1<\gamma<2$) one should expect
$\langle \delta^2 n(E) \rangle=\chi \langle n(E) \rangle$, where
$\chi = \gamma-1$ \cite{kravtsov97,detomasi19}.


A major concern for the variance method is that it relays on unfolding of
the spectrum. For a rather smooth spectra, the details of unfolding
and averaging over realizations should not 
affect the results, but for the NEE phase, where a non-smooth
spectral density is expected \cite{x2,garcia16,detomasi19}, the unfolding procedure might strongly
influence results.  A different way to study the properties of an ensemble of
spectra originating from different realizations has been recently suggested \cite{fossion13,torres17,torres18},
based on techniques originating in signal analysis. 
Given $L$ realizations of $P$ eigenvalues each, one defines a matrix
$X$ of size $L \times P$ where $X_{lp}$ is the $p$ level of the $l$-th realization. $X$ is SVD decomposed as
$X=U \Sigma V^T$, where $U$ and $V$  are $L\times L$ and $P \times P$ matrices correspondingly,
and $\Sigma$ is a {\it diagonal} matrix of size $L \times P$ and rank $r=\min(L,P)$. The
$r$ diagonal elements of $\Sigma$, denoted as $\sigma_k$ are the singular values of $X$
and may be ordered such that $\sigma_1 \geq \sigma_2 \geq \ldots \sigma_r$.
Essentially this is a non-periodic mode decomposition of the series. Defining
$\lambda_k=\sigma_k^2$ represents the fraction of the total variance in the series captured
by the mode. The lower singular values capture the global trends of
the spectra, while the higher values represent the local fluctuations. It has been postulated that
for the for higher values in the localized regime $\lambda_{k} \sim k^{-2}$ while in the extended regime
$\lambda_{k} \sim k^{-1}$ \cite{torres17,torres18}, corresponding to 1/f noise behavior \cite{relano02}.


We have calculated $\langle \delta^2 n(E) \rangle$ for large matrices of
size $N=16000,24000,32000,48000$ and corresponding $800,200,100,100$ different realizations.
For each realization $N$ eigenvalues, $\epsilon_i$, were obtained. The spectrum was then unfolded
by $\varepsilon_i=\varepsilon_{i-1}+2m(\epsilon_i-\epsilon_{i-1})/\langle \epsilon_{i+m}-\epsilon_{i-m}\rangle$
where $\langle \ldots \rangle$ is an average over realizations, and $m=6$ (other values
were used with no significant change). The center of the energy window is set at $E(j)=N/2+j\cdot 20$,
where for each realization $j=-j_{max}\ldots j_{max}$ (with $j_{max}=150,225,300,450$, i.e,
the center of the energy window is located within a range of $3/16$ of the spectra around the middle).
For each $E(j)$, the number of states in a window of width $E$ centered at $E(j)$, 
$n_j(E)$ is evaluated, then the averages $\langle n(E) \rangle$
and $\langle n^2(E) \rangle$ are taken over all positions of the center $j$ and realizations.

\begin{figure}
\includegraphics[width=8cm,height=!]{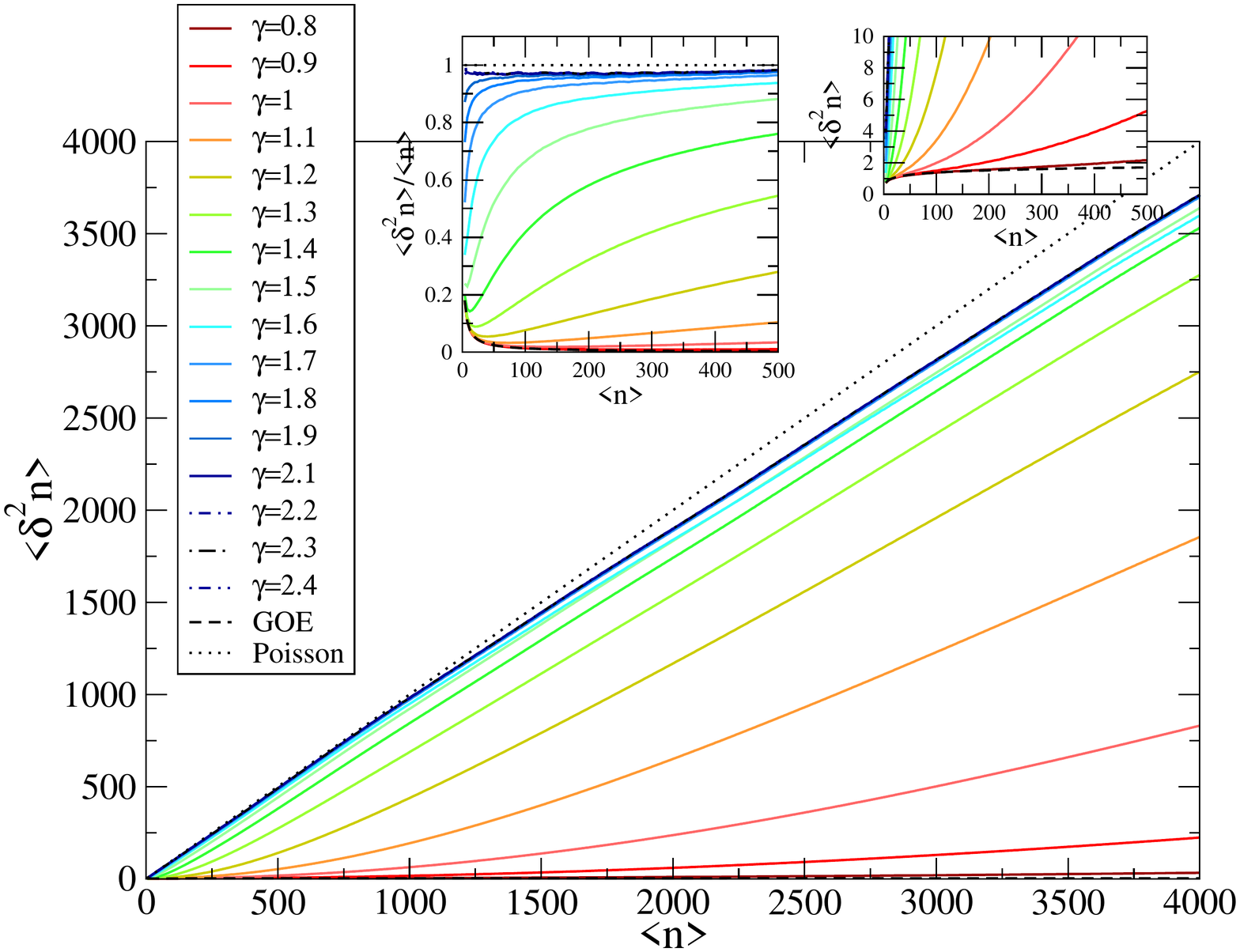}
\caption{\label{fig2}
The variance $\langle \delta^2 n(E) \rangle$ as function of  $\langle n(E) \rangle$
for the largest matrix size $N=48000$ and values of $\gamma$ between $0.8$ to
$2.4$. The Poisson and
Wigner Dyson behavior
are indicated. The top right inset zooms into the logarithmic behavior region, where deviations
from the Wigner Dyson behavior is seen even deep in the extended ergodic regime $\gamma<1$.
The middle inset the variance normalized by $n$ is plotted.
Even  in the localized regime $\gamma>2$ some deviation from
the Poisson value of $1$ is seen.
 }
\end{figure}

\begin{figure}
\includegraphics[width=8cm,height=!]{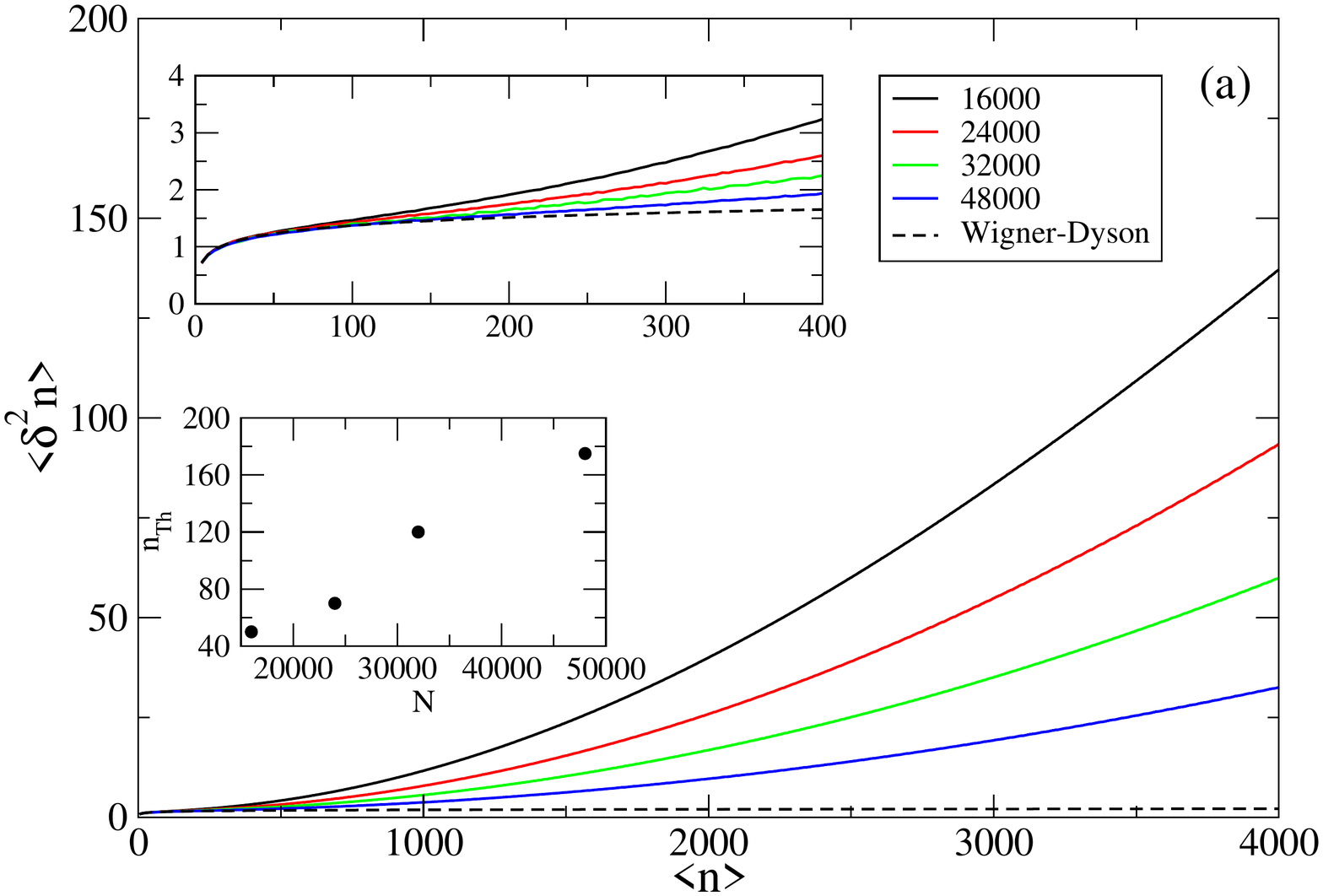}
\includegraphics[width=8cm,height=!]{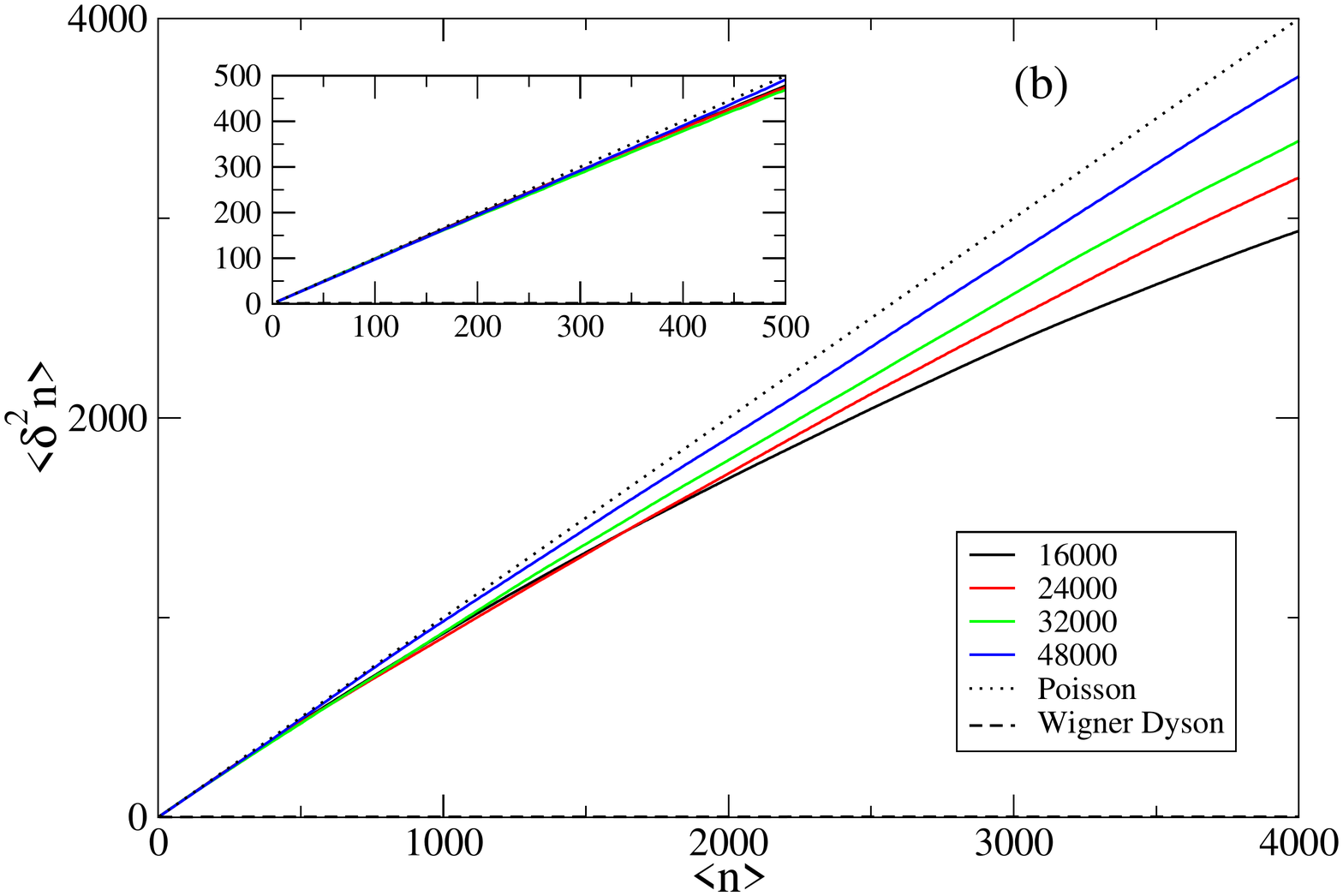}
\includegraphics[width=8cm,height=!]{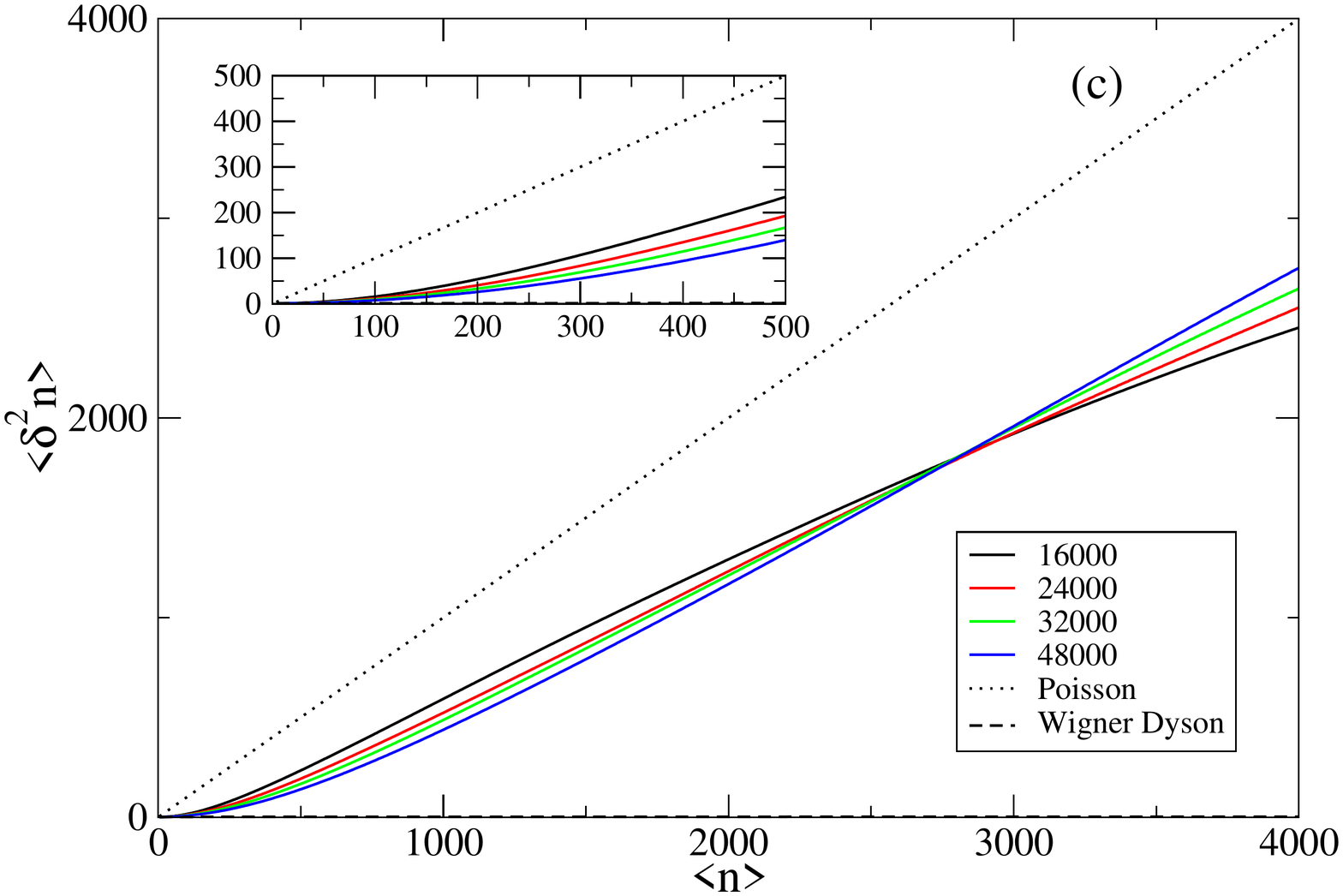}
\caption{\label{fig3}
The variance $\langle \delta^2 n(E) \rangle$ as function of  $\langle n(E) \rangle$
for (a) $\gamma=0.8$ (WD regime); (b) $\gamma=2.4$ (Poisson regime);
(c) $\gamma=1.2$ (NEE regime);
and matrix sizes between $N=16000$ and $N=48000$. 
A zoom into smaller values of $n$ is shown
in the inset.  (a) Deep in the WD regime, it is clear that at $n_{Th}$ deviations from the WD
behavior are seen. 
The dependence of $n_{Th}$ on size is presented in the lower inset in (a).
Above $n_{Th}$ 
the variance grows stronger than linear and depends on $N$.
(b) Deep in the Poisson regime. Up to $n \sim 500$ the behavior is linear as
expected from the Poisson regime. For higher $n$ deviations to weaker dependence
is seen. For larger systems the deviation appears for
larger values of $n$. (c) In the NNE regime, the behavior deviates
from WD almost immediately and follows a stronger than linear
behavior, up to a point where a weaker dependence on $n$ is seen. } 
\end{figure}

The variance $\langle \delta^2 n(E) \rangle$ as function of  $\langle n(E) \rangle$
is plotted in Fig. \ref{fig2} for the largest matrix size $N=48000$ and different
values of $\gamma$. Our main aim is to study the asymptotic
behavior of the variance at large energy windows.
Clearly as $\gamma$ increases the variance switches from
a Wigner-Dyson like behavior to a Poisson like behavior. Nevertheless, the observed behavior
raises serious doubts on our ability to give definite answers on the asymptotic behavior from
such data. Several factors compound the problem. It is clear that even for $\gamma<1$ for which
Wigner Dyson behavior ($\langle \delta^2 n(E) \rangle = (2/\pi^2)\ln(\langle n(E) \rangle)+0.44$)
is expected, this behavior is followed only up to a certain  $n_{Th}$, above which a stronger than
linear dependence is seen.
This scale $n_{Th}$ depends both on $\gamma$ (see right insert in Fig.
\ref{fig2}) and on the size of the matrix (see Fig. \ref{fig3}a). 
This is similar to the deviation seen for the Anderson model
\cite{altshuler86,altshuler88,braun95}
and in the SYK model \cite{garcia16}, and is an indication for an energy scale,
known as the Thouless energy,
$E_{Th}=\delta n_{Th}$ 
related to a time scale $t_{Th}=\hbar/E_{Th}$ indicating the typical time necessary to explore the
available system phase space system. $n_{Th}$ grows as the the system becomes less sparse
(lower $\gamma$) or larger in size.

As can be seen in Fig. \ref{fig2} for $\gamma>2$, and for different sizes in Fig. \ref{fig3}b, deep
in the Poisson regime ($\gamma>2$), the expected
Poisson behavior ($\langle \delta^2 n(E) \rangle = \langle n(E) \rangle$) is seen up to some
value of $n$ ($n \sim 1000$ for $N=48000$ and $n \sim 500$ for $N=16000$). Above this value
$\langle \delta^2 n(E) \rangle$ grows weaker than linear.  One could speculate that these
deviations are a result of  combination of finite size
effects and the unfolding which becomes less reliable at larger scales.

For the intermediate values of $1<\gamma<2$, where the NEE regime is expected,
the behavior is even messier (Fig. \ref{fig2} and Fig. \ref{fig3}c).
Almost immediately the variance starts growing much faster than
linear. As $n$ increases, the growth patters out. One might fit
a linear
behavior with a smaller than one slope, corresponding to the expected
$\langle \delta^2 n(E) \rangle=\chi \langle n(E) \rangle$ behavior.
Nevertheless, larger values $n$ remain strongly dependent on the size of the
matrix (Fig. \ref{fig3}c). Thus, it is difficult to tease out the behavior at large $n$ without ad-hoc
assumptions on the range of the fit.

\begin{figure}
\includegraphics[width=8cm,height=!]{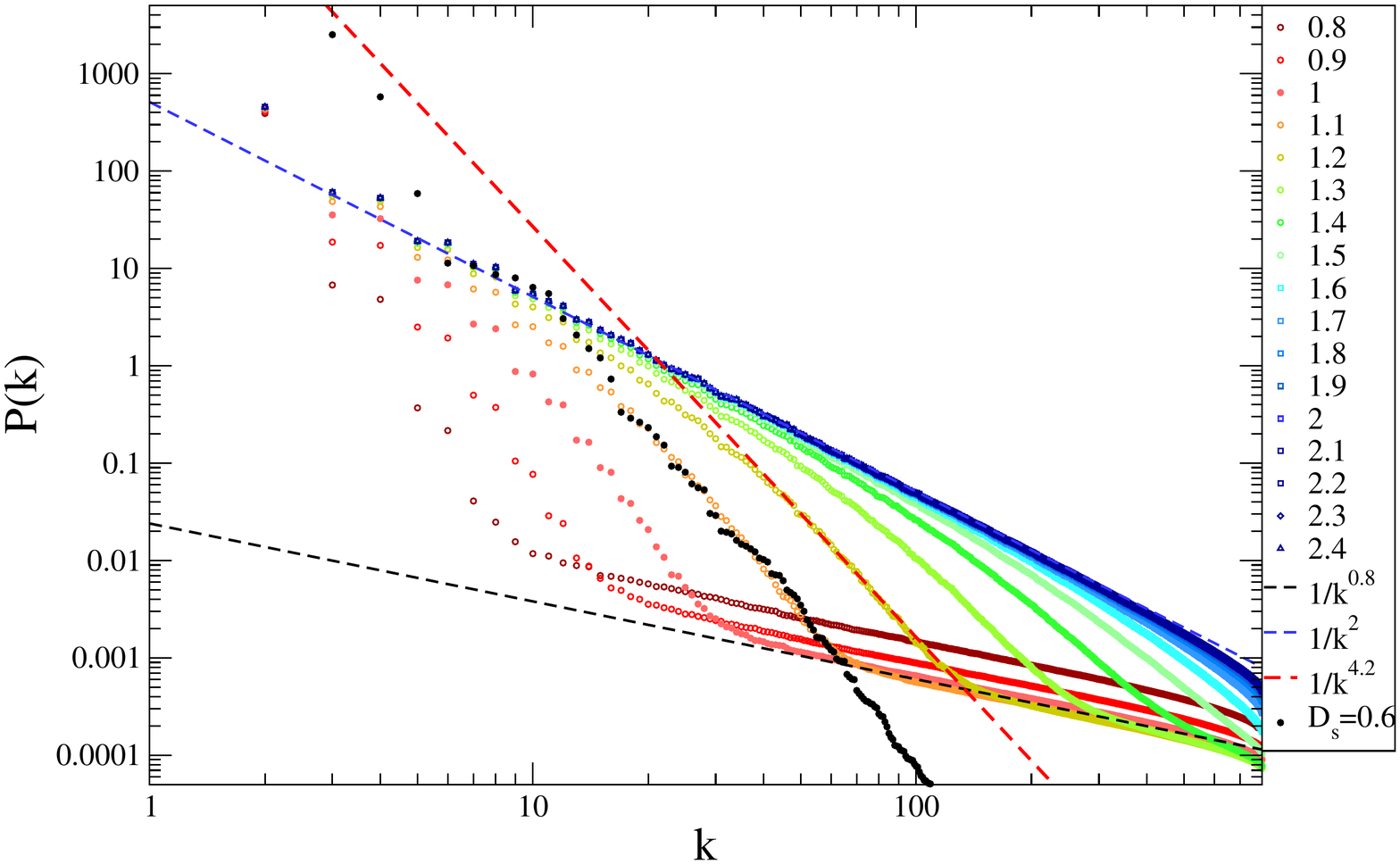}
\caption{\label{fig4}
The the scree plots of ordered
partial variances, $\lambda_k$,
for $800$ realizations of matrix size $N=16000$ and different values of $\gamma$ between $0.8$ to
$2.4$. The expected behavior for
Poisson ($1/k^{2}$), WD ($1/k^{0.8}$), and the transient behavior at $\gamma=1.2$ ($1/k^{4.2}$) are
indicated by dashed lines. A random Cantor set spectra of $N=4000$ and $1000$ realization with $D_s=0.6$
(see text) corresponds to the full black circles.
 }
\end{figure}

Due to these difficulties, we switch to the SVD  method, which does not require unfolding. 
The scree plots of the ordered
partial variances, $\lambda_k$,
for $800$ realizations of matrix size $N=16000$ and $0.8<\gamma<2.4$ is presented
in Fig.  \ref{fig4}. Large $k$ corresponds to small energy scales, 
Large energy scales depend on the overall density
of states, which is not universal, and therefore no information can be gleaned from
$k\sim 1$. For the  Poisson regime ($\gamma \ge 2$, indicated by purple symbols)
$\lambda_{k>2}$ follows the expected $k^{-2}$ behavior \cite{torres17,torres18}, 
up to deviations for large values of $k>500$.
In the WD regime ($\gamma \le 1$, indicated by reddish symbols), for $k \geq 10$,
$\lambda_{k}$ follows $k^\alpha$, with a slope $\alpha \sim 0.8$ -- different than the
expected slope  $\alpha=1$. As discussed in
the supplementary material \cite{supp}, this is the effect of a finite number of realizations, much
smaller  than the number of eigenvalues ($L \ll P$).
So for small energy scales, the
WD behavior is followed up to an energy scale, which may be identified as $E_{th}$ (corresponding
to a value $k_{Th}$) above which a much steeper decent of $\lambda_{k<k_{Th}}$  is observed. This is
in line with the number variance behavior.


The NEE regime ($1<\gamma<2$) shows intermediate  behavior between WD behavior
at large $k$ and Poisson at small values  of $k$. This general behavior is expected, since as can
be deduced from Fig. \ref{fig3}c, at very short energy scales WD behavior is expected. At values
of $\gamma>1.6$ strong deviations from Poisson are seen at large values of $k$, and for
$\gamma<1.6$ large values of $k$ show clear 
correspondence with WD. The rang of energies for which WD holds increases as $\gamma$ decreases.
For large energies (small $k$), the complementary behavior is evident, Poisson  behavior
is followed, where for larger $\gamma$ the Poisson curve is joined earlier. Thus, for large
energies  the spectra follows Poisson behavior in agreement with the expectations of Ref.
\onlinecite{kravtsov97}.

The crossover between the WD and Poisson behavior at intermediate values of $k$ is most
pronounced for $1<\gamma<1.6$. It seems that for a significant range of $k$ a definite slope
is followed,  with a slope larger than Poisson (super-Poisson), which depends on $\gamma$.
This is demonstrated for $\gamma=1.2$ where a fit to $1/k^{4.2}$ is drawn. It is evident
that a good fit in the range $40<k<120$ is obtained. A super-Poisson (or multi-fractal metal)
behavior has its origin in clustering of eigenvalues related to the mini-band structure \cite{x2,detomasi19}
for which the correlations between levels belonging to the same cluster (mini-band)  are much stronger
than between the mini-bands. In Ref. \onlinecite{x2} a random Cantor set model which mimics the expected
behavior of these mini-bands was proposed. Level spacings $\Delta$ are drawn independently
from a power-law distribution $P(\Delta>\Delta_0) \sim \Delta_0/\Delta^{1+D_s}$ (where, $\Delta_0$ is a constant,
and $D_S$ is a measure of fractality of the spectrum). As can be seen in Fig. \ref{fig4}, a
random Cantor set, drawn from $1000$ realizations of $4000$ levels each, with $D_S=0.6$, follows
quite strikingly  the behavior of $\gamma=1.1$ for intermediate values of $k$. Similarly,  decreasing
values of $D_s$ fit increasing values of $\gamma$.
This lends strong
support to the notion of a fractal  (mini-band) structure of the spectrum of the NEE
phase for intermediate energy scales.

\begin{figure}
\includegraphics[width=8cm,height=!]{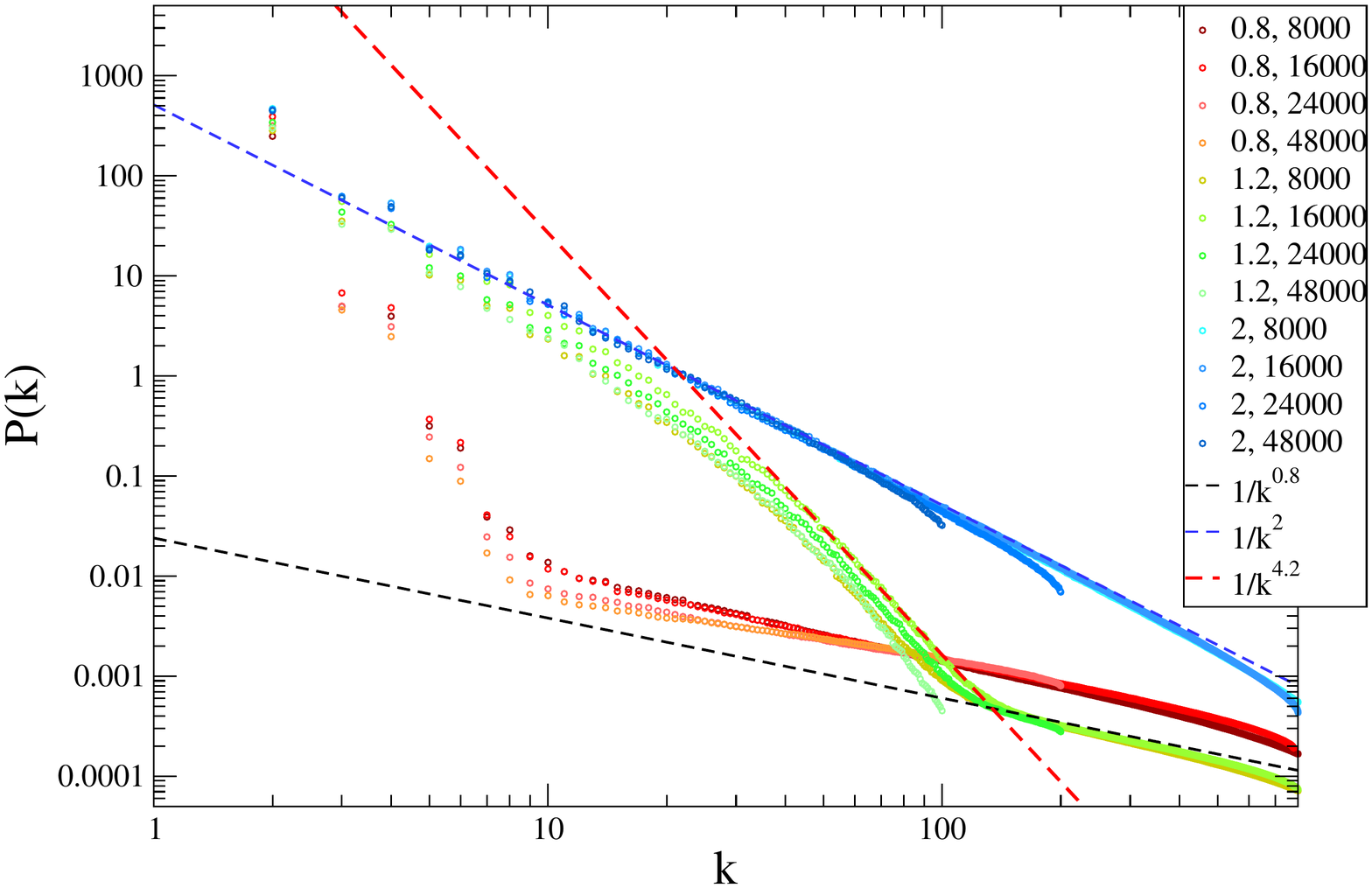}
\caption{\label{fig5}
The the scree plots of ordered
partial variances $\lambda_k$
for different matrix size $N=8000,16000,24000,480000$ for $1000,800,200,100$ realizations correspondingly,
and three different values of $\gamma=0.8,1.2,2$.
The curves for different matrix sizes where shifted by multiplying them by a constant in order that they
will overlap with the  curves for $N=16000$. The influence of the number of realizations is depicted by
calculating $\lambda_k$ for $N=16000$, with only $400$ and $200$ realizations for  $\gamma=1.2$ (black
symbols).
}
\end{figure}

As for the number variance one may wonder how sensitive is the SVD method to
finite size effects. In Fig. \ref{fig5}, we examine the dependence of the scree plot slopes
on matrix sizes $N=8000,16000,24000,480000$ with  $1000,800,200,100$ realizations
for WD ($\gamma=0.8$), Poisson ($\gamma=2$) and the NEE ($\gamma=1.2$)
regime. Since the value of $\lambda_k$ depends on size, for comparison we multiplied the
curves by a constant to shift them one on top of the other for the same $\gamma$.
In all cases a similar behavior of $\lambda_k$ is seen for all sizes.
The same holds for changing the number of realizations. Of course the maximum $k$ is reduced
but the overall behavior remains, as can be seen from Fig. \ref{fig5}
where for $N=16000$ and $\gamma=1.2$, with $800, 400$ and $200$ realizations.  

Thus, singular value decomposition reveals robust super-Poissonian behavior for intermediate energy scales
of the NEE phase for the parameter range $1.6<\gamma<2$. For $1<\gamma<1.6$, it is hard to observe the super-Poissonian
regime since it is pushed to smaller energy scales, i.e., larger $k$. Moreover, the absolute
slope decreases, and thus the deviation from WD is less pronounced. Studying this regime will require a much larger
number of realization than is available for this study. Demonstration of super-Poissonian behavior of the
SVD analysis of energy spectra of other systems which are expected to show NEE behavior, for example,
disordered Josephson junctions array \cite{x2} and
granular SYK matter \cite{altland19} may turn out very illuminating.



\begin{thebibliography}{99}


\bibitem{anderson58} P.W. Anderson, Phys. Rev., {\bf 109}, 1492 (1958).

\bibitem{lee85} P. A. Lee and T. V. Ramakrishnan, Rev. Mod. Phys. {\bf 57}, 287 (1985).

\bibitem{kramer93} B. Kramer A. MacKinnon, Rep. Prog. Phys. {\bf 56}, 1469 (1993).

\bibitem{aoki83} H. Aoki, J. Phys. C {\bf 16}, L205 (1983).

\bibitem{shklovskii93} B. Shklovskii, B. Shapiro, B. R. Sears, P. Lambrianides and H. B. Shore, Phys. Rev. B. {\bf 47}, 11487
  (1993).

\bibitem{ghur98}
T. Guhr, A. Muller-Groeling, H. A. Weidenmuller,  Phys. Rep. {\bf 299}, 190 (1998).

\bibitem{mirlin00}
A.D. Mirlin, Phys. Rep. {\bf 326}, 259 (2000).

\bibitem{evers08}
R. Evers and A.D. Mirlin, Rev.  Mod. Phys. {\bf 80}, 1355 (2008).

\bibitem{kravtsov94} V. E. Kravtsov, I. V. Lerner, B. L. Altshuler and A. G. Aronov, Phys. Rev. Lert {\bf 72}, 888 (1994).

\bibitem{evangelou94} S. N. Evangelou, Phys. Rev. B {\bf 49}, 16805 (1994).

\bibitem{aronov94} A. G. Aronov, V. E. Kravtsov, and I. V. Lerner, Pis'ma Zh. Eksp.
  Teor. Fiz. {\bf 59}, 40 (1994) [JETP Lett. {\bf 59}, 39 (1994)].

\bibitem{zharekeshev97} I. Kh. Zharekeshev and B. Kramer, Phys. Rev. Lett. {\bf 79}, 717 (1997).
  
\bibitem{altshuler86} B. Altshuler and B. Shklovskii, Sov. Phys. JETP [Zh. Eksp. Teor. Fiz. 91,220]
  {\bf 64}, 127 (1986).

\bibitem{altshuler88} B. Altshuler, I. Zarekeshev, S. Kotochigova, and B. Shklovskii, Sov. Phys. JETP [Zh. Eksp.Teor. Fiz. 94, 343] {\bf 67}, 625 (1988).

\bibitem{aronov95} A. G. Aronov and A. D. Mirlin
Phys. Rev. B {\bf 51}, 6131(R) (1995).

\bibitem{x1} M. Pino, L. B. Ioffe and B. L. Altshuler, PNAS {\bf 113},
536 (2016).

\bibitem{x2} M. Pino, V. Kravtsov, B. Altshuler and L. Ioffe, Phys. Rev. B {\bf 96}, 214205 (2017).

\bibitem{x3} T. Mithun, Y. Kati, C. Danieli and S. Flach, Phys. Rev. Lett. {\bf 120}, 184101 (2018).

\bibitem{x4} M. Thudiyangal, C. Danieli, Y. Kati and S. Flach, Phys. Rev. Lett. {\bf 122}, 054102 (2019).

\bibitem{x5} E. J. Torres-Herrera and L. F. Santos, Annalen der Physik {\bf 529}, 1600284 (2017).

\bibitem{x6} R. Berkovits, Annalen der Physik {\bf 529}, 1700042 (2017).

\bibitem{x7} J. Lindinger, A. Buchleitner and A. Rodr\'ıguez Phys. Rev. Lett. {\bf 122}, 106603 (2019).
t
\bibitem{x8} S. Roy,I. Khaymovich, A. Das and R. Moessner, SciPost Physics {\bf 4} 025 (2018).

\bibitem{x9} L. Faoro, M. Feige\'lman and L. Ioffe, arXiv:1812.06016 (2018).

\bibitem{x10} K. Kechedzhi, V. Smelyanskiy, J. R. McClean, V. S. Denchev, M. Mohseni, S. Isakov, S. Boixo, B. Altshuler and H. Neven arXiv:1807.04792 (2018).

\bibitem{sachdev93} S. Sachdev and J. Ye, Phys. Rev. Lett. {\bf 70}, 3339 (1993).
  
\bibitem{maldacena99} J. Maldacena, International journal of theoretical physics {\bf 38}, 1113 (1999).
  
\bibitem{maldacena16} J. Maldacena, S. H. Shenker, and D. Stanford, J. High Energ. Phys. {\bf 08} 106  (2016). 

\bibitem{micklitz19}  T. Micklitz, F. Monteiro and A. Altland, Phys. Rev. Lett. {\bf 123}, 125701 (2019).

\bibitem{garcia16} A. M. Garc\'ia-Garc\'ia and J. J. M. Verbaarschot Phys. Rev. D 94, 126010 (2016)

\bibitem{rosenzweig60} N. Rosenzweig and C. E. Porter, Phys. Rev. B {\bf 120}, 1698 (1960).
  
\bibitem{kravtsov15} V. E. Kravtsov, I. M. Khaymovich, E. Cuevas, and M. Amini,
New J. Phys. {\bf 17} (2015).

\bibitem{monthus17}  C. Monthus, J. Phys. A: Math. Theor. {\bf 50}, 295101 (2017).

\bibitem{kravtsov18}  V. Kravtsov, B. Altshuler and L. Ioffe, Ann. Phys. {\bf 389}, 148 (2018).

\bibitem{bogomolny18}  E. Bogomolny and M. Sieber, Phys. Rev. E {\bf 98}, 032139 (2018).

\bibitem{nosov19} P. Nosov, I. M. Khaymovich and V. E. Kravtsov, Phys. Rev. B {\bf 99}, 104203 (2019).

\bibitem{pino19} M. Pino, J. Tabanera and P. Serna, arXiv:1904.02716.
  
\bibitem{detomasi19} G. de Tomasi, M. Amini, S. Bera, I. M. Khaymovich, and V. E.
  Kravtsov, SciPost Phys. {\bf 6}, 14 (2019).

\bibitem{mehta91} M. L. Mehta, {\it Random matrices} (Acad. Press, New York,
1991), 2nd ed.

\bibitem{fossion13} R. Fossion, G. Torres-Vargas and J. C. L\'opez-Vieyra,
Phys. Rev. E, {\bf 88}, 060902(R) (2013).

\bibitem{torres17}
G. Torres-Vargas, R. Fossion, C. Tapia-Ignacio and J. C.
L\'opez-Vieyra, Phys. Rev. E, {\bf 96}, 012110 (2017).

\bibitem{torres18}
G. Torres-Vargas, J. A. M\'endez-Berm´udez, J. C. L\'opezVieyra and R. Fossion, Phys. Rev. E, {\bf 98}, 022110 (2018).

\bibitem{relano02} A. Rela\~{n}o, J. M. G. G\'{o}mez, R. A. Molina, J. Retamosa, and E. Faleiro, Phys. Rev.  Lett. {\bf 89},
  244102 (2002).

\bibitem{braun95} D. Braun and G. Montambaux, Phys. Rev. B {\bf 52}, 13903 (1995).

\bibitem{pandey95} A. Pandey, Chaos Solitons Fractals {\bf 5}, 1275 (1995).

\bibitem{brezin96} E. Brezin and S. Hikami, Nucl. Phys. B {\bf 479}, 697 (1996).

\bibitem{guhr96} T. Guhr, Ann. Phys. {\bf 250}, 145 (1996).

\bibitem{altland97} A. Altland, M. Janssen, and B. Shapiro, Phys. Rev. E {\bf 56}, 1471 (1997).

\bibitem{kunz98} H. Kunz and B. Shapiro, Phys. Rev. E {\bf 58}, 400 (1998).

\bibitem{facoetti16} D. Facoetti, P. Vivo, and G. Biroli, Europhys. Lett. {\bf 115}, 47003 (2016).

\bibitem{supp}
  See Supplemental Material at [URL] for nearest level  spacing transition at
  $\gamma=2$ and SVD scree plots of lageg $1/f$ noise series.
  
\bibitem{kravtsov97} V. E. Kravtsov, K. A. Muttalib, Phys. Rev. Lett. {\bf 79}, 1913
  (1997).

\bibitem{altland19}  A. Altland, D. Bagrets, and A. Kamenev, Phys. Rev. Lett. {\bf 123}, 106601 (2019).










\end{thebibliography}
\end{document}


\title{Supplemental material for ``On super-Poissonian behavior of the Rosenzweig-Porter model in the non-ergodic extended regime''}
\author{Richard Berkovits}
\affiliation{Department of Physics, Jack and Pearl Resnick Institute, Bar-Ilan University, Ramat-Gan 52900, Israel}


\maketitle

\section{Introduction}
In this supplemental material
the transition from the extended (NEE) to the localized regime as manifested in the finite size
scaling of the nearest level spacing is studied in the first section, while in the next section the
SVD scree plot for a large simulated $1/f$ noise series is discussed.

\section{Nearest level spacing transition at $\gamma=2$}

\begin{figure}
\includegraphics[width=8cm,height=!]{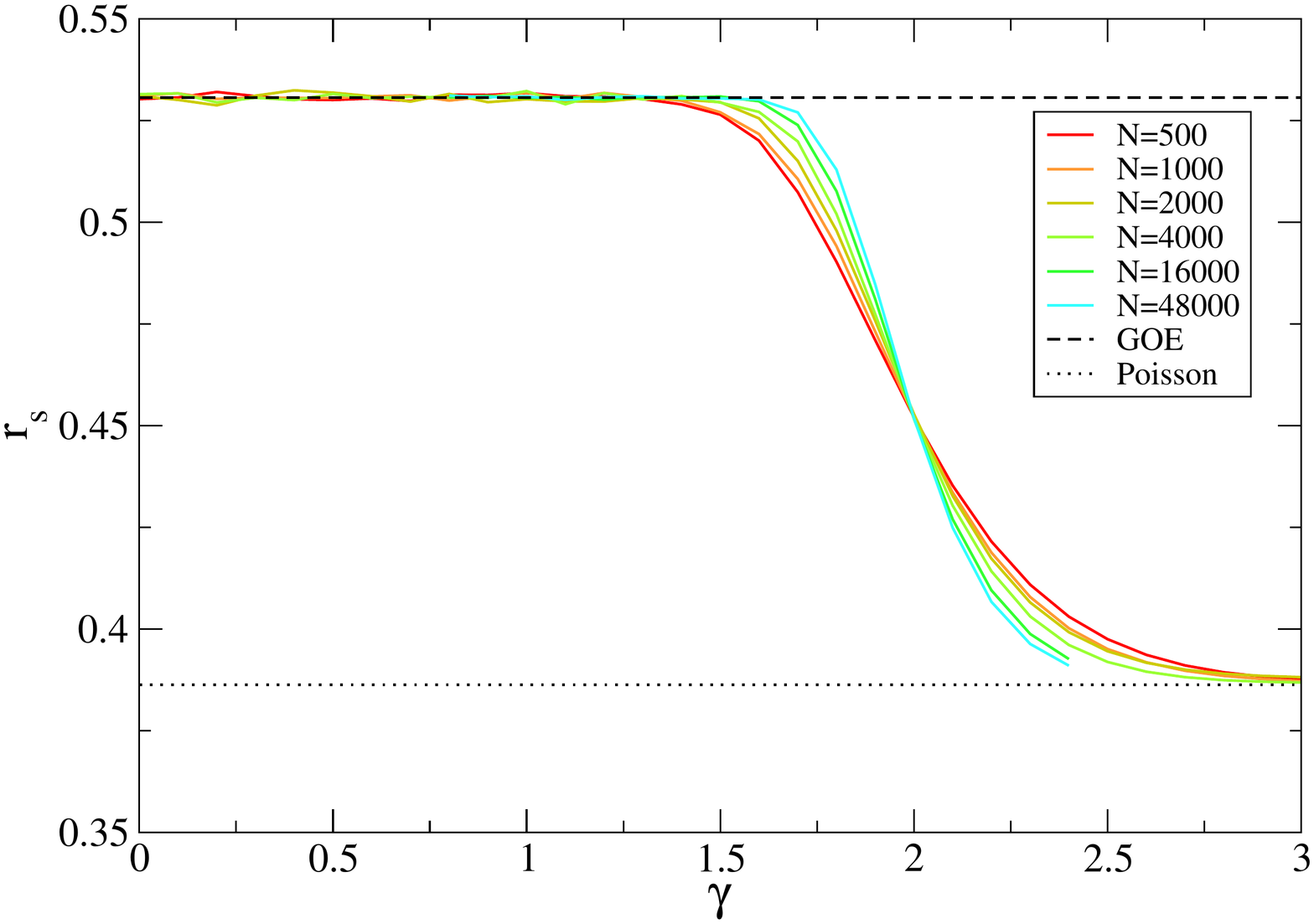}
\includegraphics[width=8cm,height=!]{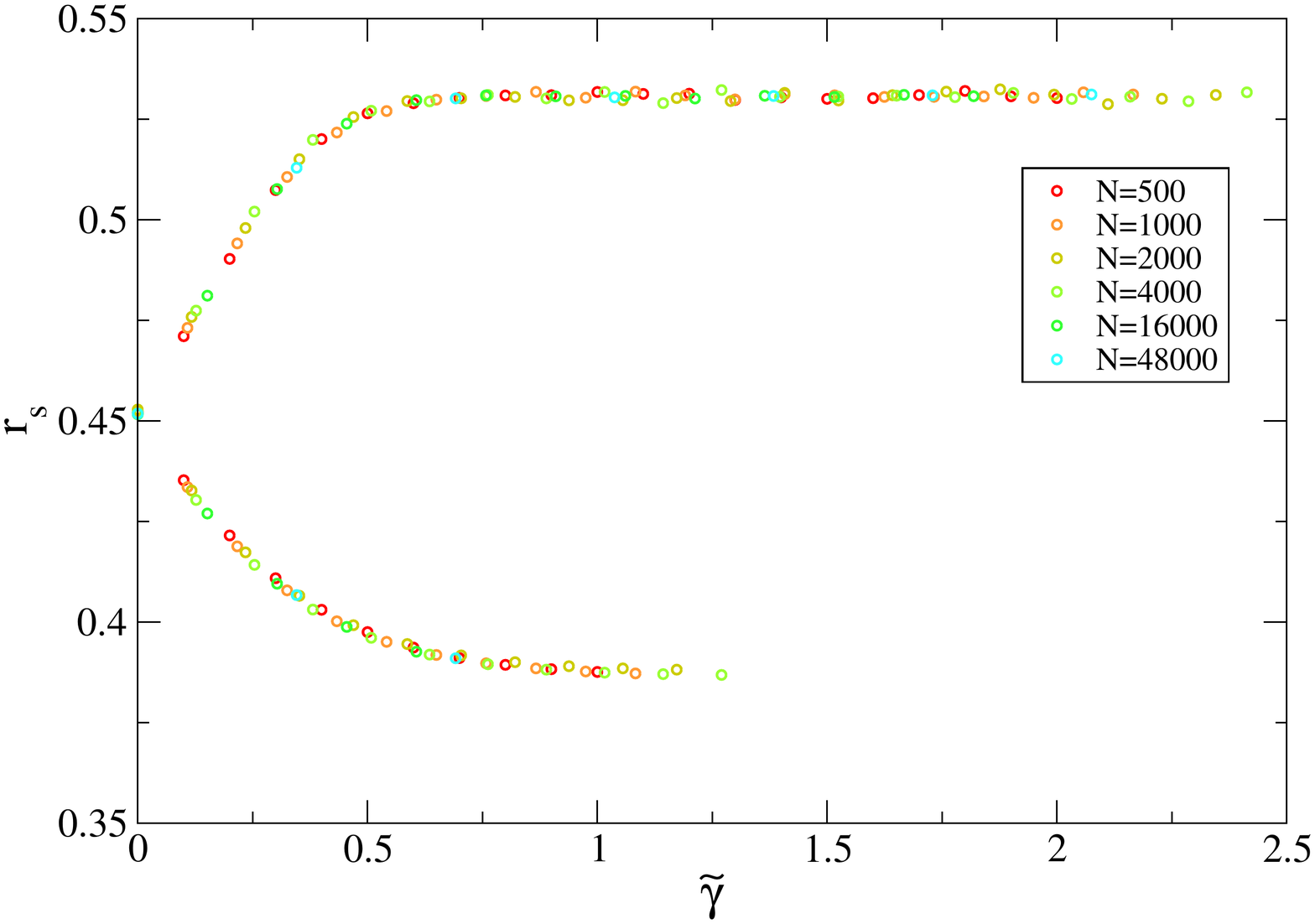}
\caption{\label{fig1}
  The nearest neighbor level spacing statistics as manifested in the
  behavior of the ratio statistics $r(\gamma)$ for different values of $\gamma$ and
  matrix size $N$. (a) The finite size dependence of  $r_s(\gamma)$. All curves cross
  at the critical value $\gamma_c=2$ and exchange their order as they cross this point. This
  is a clear manifestation of a second order phase transition. (b) Finite size scaling. All
  curves neatly fall on the two branches above and below the transition as function of
  $\tilde{\gamma}$ defined in the text.
}
\end{figure}

For generalized Rosenzweig-Porter (GRP) one expects a transition from localized to extended behavior
at $\gamma=2$ \cite{kravtsov15,detomasi19,pandey95,brezin96,guhr96,altland97,kunz98,facoetti16}.
This transition is manifested in the energy spectrum of the system in a transition
in the nearest neighbor level distribution between Poisson behavior for $\gamma>2$ to
Wigner-Dyson like repulsion at $\gamma<2$.
The finite size scaling of this behavior
is clearly seen by the ratio statistics, defined as:
\begin{eqnarray} \label{ratio}
r_s &=& \langle \min (r_n,r_n^{-1}) \rangle,
\\ \nonumber
r_n &=& \frac {E_n-E_{n-1}}{E_{n+1}-E_{n}},
\end{eqnarray}
where $E_n$ is the n-th eigenvalue of the matrix and $\langle \ldots \rangle$ is
an average over different matrices and a range of eigenvalues.
For the Poisson distribution one expects $r_s=2\ln(2) - 1 \sim 0.3863$,
while for the WD  distribution $r_s \sim  0.5307$ \cite{atas13}. Finite size
scaling assumes that for $\gamma>2$ the value of $r_s$ approaches the Poisson value as the matrix size grows,
while for $\gamma<2$, $r$ approaches the WD value. At the transition ($\gamma=2$) $r$ should be
independent of the matrix size. Thus, as one plot $r_s(\gamma)$ for larger values of $N$
the curve becomes more step-like and all the curves are expected to cross at the same point.
This has been seen in Ref. \cite{pino19} and is reproduced here for larger matrix sizes. In
Fig. \ref{fig1}a we plot $r(\gamma)$ for
$N=500,1000,2000,4000,16000,48000$ with $12800,6400,3200,1600,800,100$ different
matrices. The curves can be scaled
using $\tilde{\gamma} = |\gamma-\gamma_c|(N/N_0)^{\beta}$, with $\beta=0.12$, $\gamma_c=2$ and $N_0=500$ was
chosen as the smallest matrix size. As can be seen in Fig. \ref{fig1}b this scaling works
well.

Thus, for the low energy scale one finds very clear evidence for the transition between
localized states and extended ones at $\gamma_=2$. 

\section{SVD scree plot of large $1/f$ noise series}

\begin{figure}
\includegraphics[width=8cm,height=!]{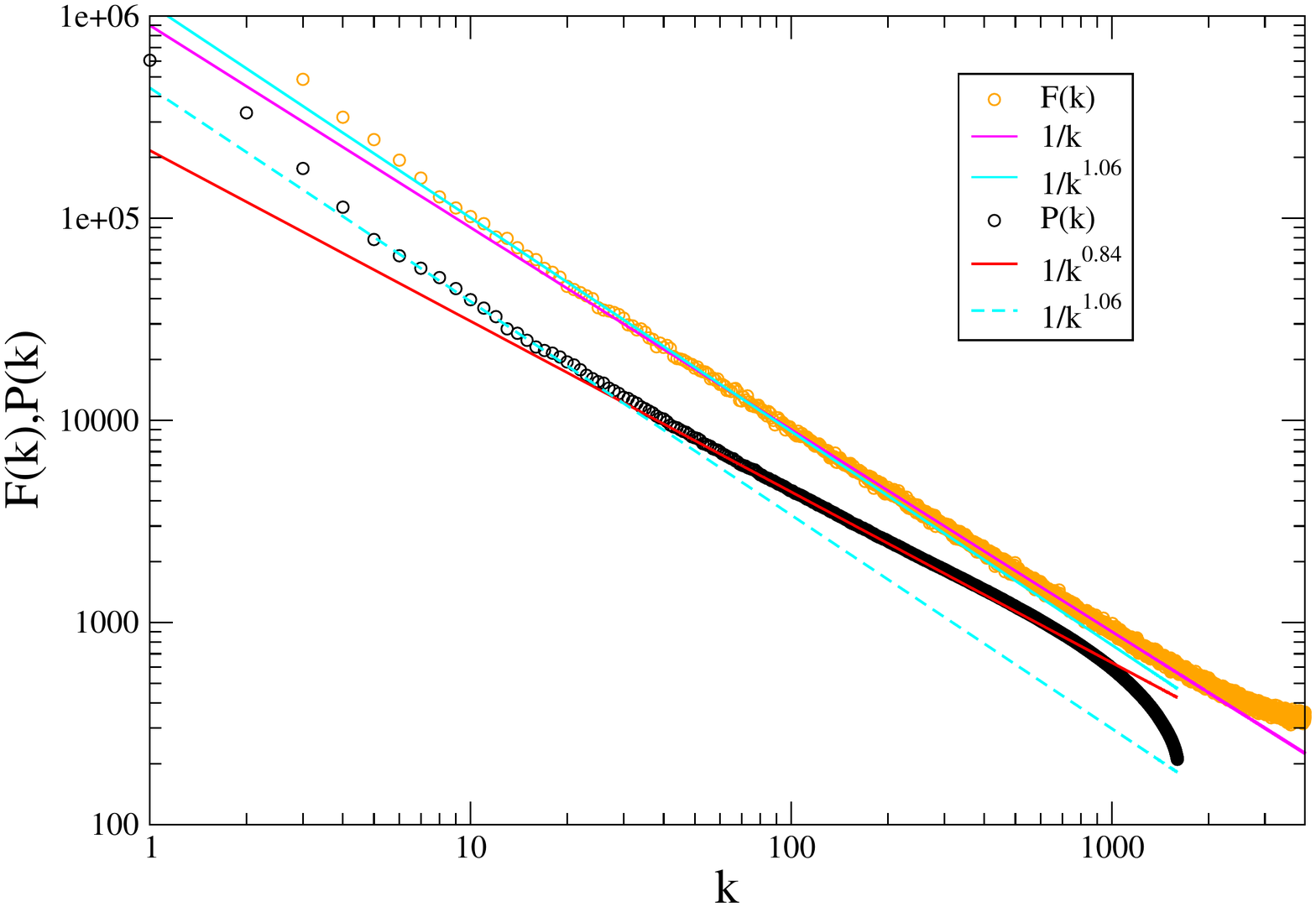}
\caption{\label{fig2}
  Power spectrum ($F(k)$ ,orange circles) and
  Scree plot of the SVD ($P(k)$, black circles)
  of an ensemble of numerically generated 1/f noise sequences of
  length $8192$ for $1600$ different realizations. 
  Attempts to fit to different slopes are shown.
}
\end{figure}

By definition a $1/f$ noise series is a sequence with a power spectrum of $1/f$. Such a series may be numerically
generated by different methods. Here we apply the method of Kasdin \cite{kasdin95}, using a recursive filter.
We generate $L=1600$ realizations with $P=8192$ values each ($\{X^{l=1,L}_{p=1,P} \}$), and perform a discreet fast
Fourier transform,
$f^l(k)=(1/\sqrt{P})\sum_p  X^l_p\exp(-2\pi i k p/P)$, and then average the power spectrum over all
realizations, $F(k)=\sum_l |f^l(k)|^2/L$ which is shown in Fig. \ref{fig2}. The SVD of the matrix $X$ is
extract as described in the main text and the
the scree plot $P(k)$, can also be seen in Fig. \ref{fig2}.

The power spectrum shows the expected $1/k$ for the range $50<k<2000$. For $k<50$, 
$F(k)$ takes a somewhat larger slope, $1/k^{1.06}$. The SVD Scree plot is
expected to follow the power spectrum slope \cite{fossion13,torres17,torres18}.
Indeed, for $k<50$,  
$P(k)$ follows with the same slope, $1/k^{1.06}$,  while 
for, $50<k<1000$  $P(k)$ corresponds to a different slope $\sim 1/k^{0.84}$.
It is important to note that while for the Fourier transform results in $0<k<P/2$
amplitudes
and the number of realizations $L$ determines only the quality of averaging,
for the SVD one is limited to $L$ eigenvalues. Therefore, if $P \ll L$,
SVD will show finite number of realization  effects for a smaller value of $k$,
resulting in the fact that although the Fourier transform shows a $1/k$ slope,
the SVD results show a gentler slope. This trend is seen also in the main text,
where although one expects the SVD in the Wigner-Dyson regime to show a $1/k$ slope,
it shows a slope closer to $\sim 1/k^{0.8}$.